\documentclass[useAMS]{mn2e} 
 
\usepackage{graphicx} 
\usepackage{txfonts} 
\newcommand\aj{AJ} 
\newcommand\apj{ApJ} 
\newcommand\apjs{ApJS}       
 
\newcommand\aap{A\&A} 
\newcommand\mnras{MNRAS} 
\newcommand\apjl{ApJ} 
\newcommand\pasp{PASP} 
\newcommand\nat{Nature}

\newcommand\aapr{ARA\&A}
 
\title[]{Multiple stellar populations in Magellanic Cloud clusters. V. The split main sequence of the young cluster NGC\,1866
\thanks{
    Based on observations with the NASA/ESA Hubble Space Telescope,
    obtained at the Space Telescope Science Institute, which is
    operated by AURA, Inc., under NASA contract NAS 5-26555, under
    GO-14204.
}}  
\author[A.\,P.\, Milone et al.] 
       {A.\,P.\,Milone$^{1}$,
         A.\,F.\,Marino$^{1}$,
         F.\,D'Antona$^{2}$,
         L.\,R.\,Bedin$^{3}$,
         G.\,Piotto$^{3,4}$,
         H.\,Jerjen$^{1}$, \newauthor
         J.\,Anderson$^{5}$,
         A.\,Dotter$^{6}$,
         M.\,Di Criscienzo$^{2}$,
         E.\,P.\,Lagioia$^{7,8}$
\\ 
$^{1}$Research School of Astronomy \& Astrophysics, Australian National University, Canberra, ACT 2611, Australia \\
         $^{2}$Istituto Nazionale di Astrofisica - Osservatorio Astronomico di Roma, Via Frascati 33, I-00040 Monteporzio Catone, Roma, Italy\\
$^{3}$Istituto Nazionale di Astrofisica - Osservatorio Astronomico di Padova, Vicolo dell'Osservatorio 5, Padova, IT-35122\\
$^{4}$Dipartimento di Fisica e Astronomia ``Galileo Galilei'', Univ. di Padova, Vicolo dell'Osservatorio 3, Padova, IT-35122\\
$^{5}$Space Telescope Science Institute, 3800 San Martin Drive, Baltimore,  MD 21218, USA\\
$^{6}$ Department of Astronomy, Harvard University, Cambridge, MA, USA \\
$^{7}$Instituto de Astrof\`isica de Canarias, E-38200 La Laguna, Tenerife, Canary Islands, Spain\\
$^{8}$Department of Astrophysics, University of La Laguna, E-38200 La Laguna, Tenerife, Canary Islands, Spain\\
} 
\begin{document} 
\date{Draft Version Oct, 21, 2016} 
 
\pagerange{\pageref{firstpage}--\pageref{lastpage}} \pubyear{2016} 
 
\maketitle 
\label{firstpage} 
 
\begin{abstract}
  One of the most unexpected results in the field of stellar populations of the last few years, is the discovery that some Magellanic-Cloud globular clusters younger than $\sim$400 Myr, exhibit bimodal main sequences (MSs) in their color-magnitude diagrams (CMDs). 
  Moreover, these young clusters host an extended main sequence turn off (eMSTO) in close analogy with what is observed in most $\sim$1-2 Gyr old clusters of both Magellanic Clouds.  

  We use high-precision {\it Hubble Space Telescope} photometry to study the young star cluster NGC\,1866 in the Large Magellanic Cloud. 
  We discover an eMSTO and a split MS. The analysis of the CMD reveals that (i) the blue MS is the less populous one, hosting about one-third of the total number of MS stars; (ii) red-MS stars are more centrally concentrated than blue-MS stars; (iii) the fraction of blue-MS stars with respect to the total number of MS stars drops by a factor of $\sim$2 in the upper MS with $m_{\rm F814W} \lesssim 19.7$.  
  
  The comparison between the observed CMDs and stellar models reveals that the observations are consistent with $\sim$200 Myr old highly-rotating stars on the red-MS, with rotation close to critical value, plus a non-rotating stellar population spanning an age interval between $\sim$140 and 220 Myr, on the blue-MS.

 Noticeable, neither stellar populations with different ages only, nor coeval stellar models with different rotation rates, properly reproduce the observed split MS and eMSTO. We discuss these results in the context of the eMSTO and multiple MS phenomenon.
\end{abstract} 
 
 
\section{Introduction}\label{sec:intro} 
Nearly all the old Globular Clusters (GCs) host multiple stellar populations with typical photometric and spectroscopic features (Gratton et al.\,2004; Piotto et al.\,2015; Marino et al.\,2015; Milone et al.\,2016a and references therein).
 The formation of the multiple stellar populations in the early Universe is one of the main open issues of stellar astrophysics and could play an important role in the assembly of the Galaxy (e.g.\,Gratton et al.\,2012; Renzini et al.\,2015).

In this context, the discovery that most intermediate-age star clusters in the Large and Small Magellanic Clouds (LMC, SMC) exhibit a multimodal or extended main-sequence turn off (eMSTO, Bertelli et al.\,2003; Mackey \& Broby Nielsen 2007; Glatt et al.\,2009; Milone et al.\,2009), and in some cases dual red clumps (Girardi et al.\,2009) has been one of the most-intriguing discoveries of the last decade in the field of stellar populations. Indeed it has been suggested that clusters with eMSTOs are the younger counterparts of the old GCs with multiple populations (e.g.\,Mackey et al.\,2008; Keller et al.\,2011).

The origin of the eMSTO has been widely investigated but a solution is still missing. A possible interpretation is that the eMSTO is due to multiple stellar populations with difference in age of about 100-700 Myr (e.g.\,Goudfrooij et al.\,2011, 2014; Li et al.\,2014) and that the intermediate-age clusters have experienced a prolonged star-formation episode in close analogy with old GCs (Conroy et al.\,2011). As an alternative, the eMSTO is due to coeval multiple populations with different rotation rates (Bastian \& De Mink 2009; D'Antona et al.\,2015) or to interacting binaries (Yang et al.\,2011, 2013).

Recent papers, based on {\it Hubble Space Telescope\,} ({\it HST\,}) photometry, have shown that the $\sim$300-Myr old cluster NGC\,1856 and the $\sim$100-Myr clusters NGC\,1844 and NGC\,1755 exhibit a very complex color-magnitude diagram (CMD) including split MS and eMSTO (Milone et al.\,2013, 2015, 2016b; Correnti et al.\,2015).
These findings have made a clear case that the once-thought simple MC young clusters host multiple stellar populations and that the eMSTO is not a peculiarity of the $\sim$1-2 Gyr star clusters. 

The presence of multiple populations in young clusters has opened a new window of opportunity to investigate the eMSTO and has provided additional constraints to discriminate among the different scenarios.
In this paper we investigate multiple stellar populations in the $\sim 200$-Myr old cluster NGC\,1866 by using {\it HST} images. 
 The paper is organized as follows. Section~\ref{sec:data} describes the dataset and the data analysis. In Sections~\ref{sec:cmd} and~\ref{sec:ms} we present the CMD of NGC\,1866 and investigate the cluster's double MS while in Section~\ref{sec:teo} we compare the observed CMD with theoretical models. A summary and discussion follow in Section~\ref{sec:discussion}. 

\section{Data and data analysis} \label{sec:data} 
\begin{centering} 
\begin{figure*} 
 \includegraphics[width=8.5cm]{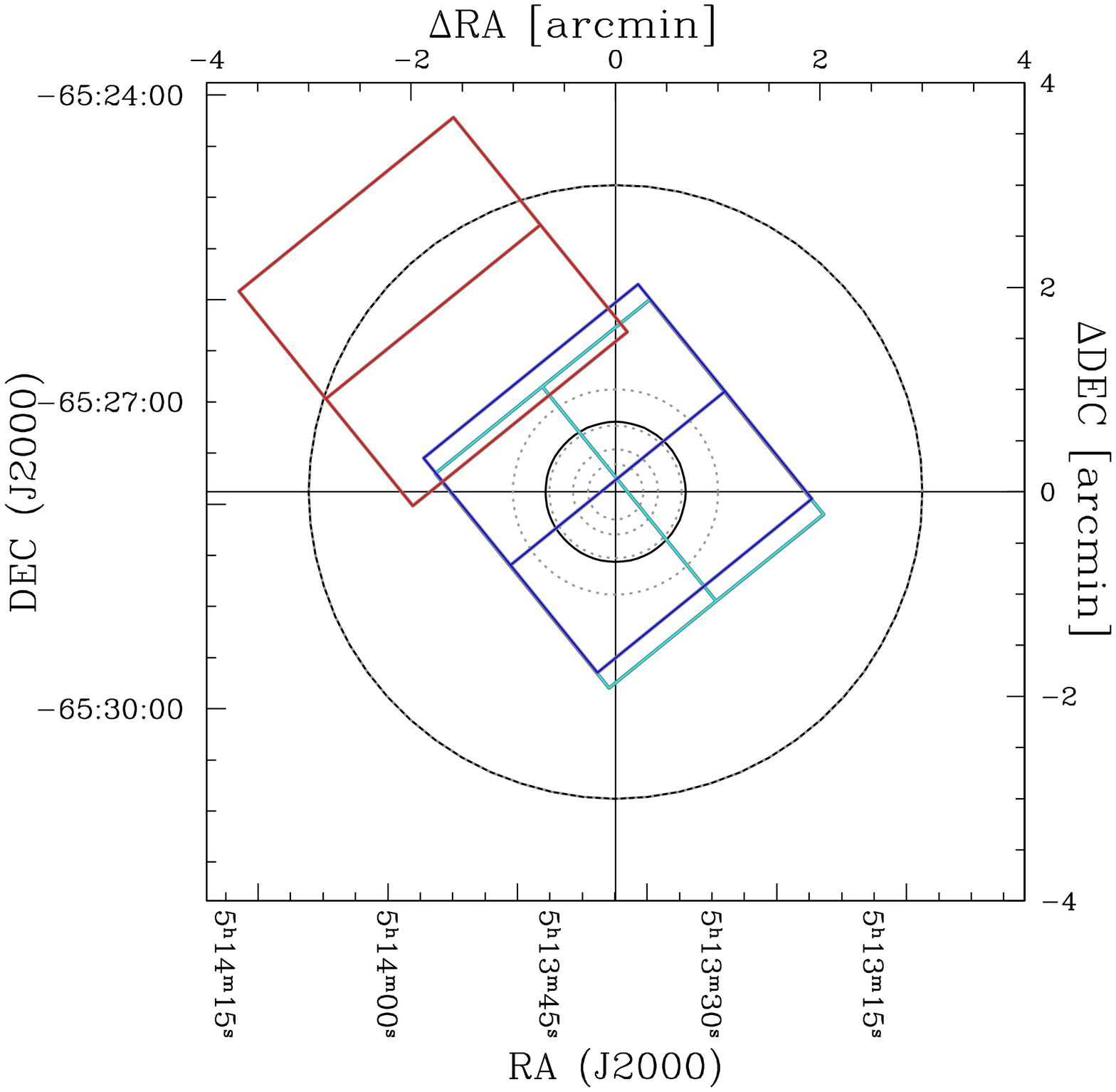} 
 \includegraphics[width=8.5cm]{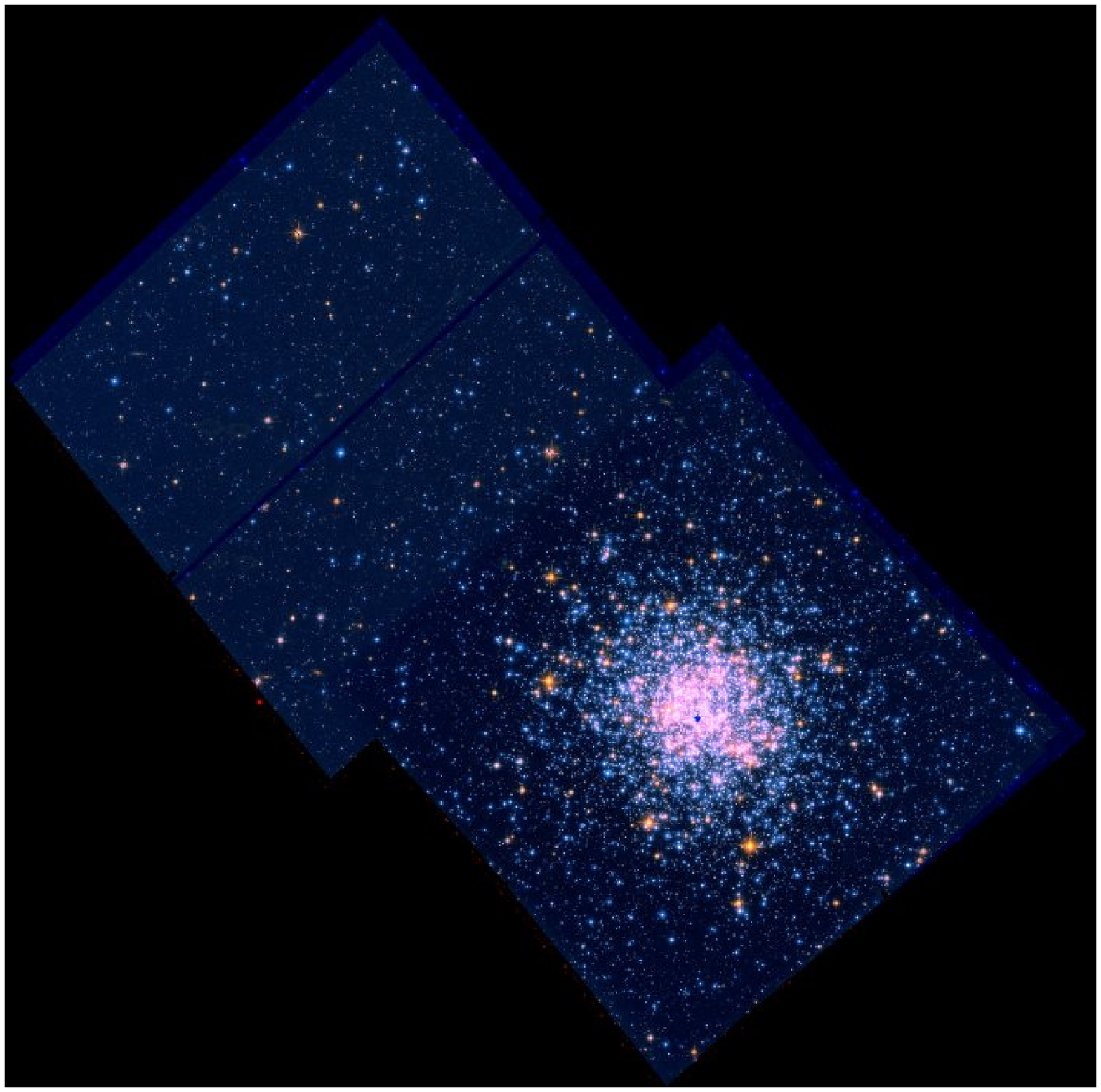} 
 \caption{\textit{Left panel:} Footprints of the UVIS/WFC3 images used in this paper. Blue, cyan, and red colors refer to images collected during different visits. The inner solid circle has a radius of 41 arcsec, corresponding to the projected cluster half-light radius, and delimits the cluster field. Reference-field stars are located outside the outer solid circle.  
   The five dotted circles indicate the regions used to study the radial distribution of stellar populations in NGC\,1866. \textit{Right panel:} Thrichromatic image of the analyzed field of view. } 
 \label{fig:footprint} 
\end{figure*} 
\end{centering} 

The dataset that we have used to investigate multiple stellar populations in NGC\,1866 has been collected through the Ultraviolet and Visual Channel of the Wide Field Camera 3 (UVIS/WFC3) of  {\it HST}. The footprints of these images are shown in the left panel of Figure~\ref{fig:footprint} where the different colors indicate images taken during three distinct visits on March, 1 (blue), May, 31 (red), and June, 1, 2016 (cyan).  Each visit includes 2$\times$711s images collected through the F336W filter and 90s$+$678s images collected through the F814W filter.
The inner and outer black-continuous circles shown in the left panel of Figure~\ref{fig:footprint} have radius of 41 (equivalent to the projected half-light radius of NGC\,1866, Mc Laughlin \& Van der Marel\,2005) and 180 arcsec, respectively.  The region within the inner circle is mainly populated by cluster members and will be designed as cluster field hereafter. In contrast, the region outside the outer circle mostly contains field stars and is called reference field.
  To determine the radius of the outer circle we have calculated the number of stars with $m_{\rm F814W}<22.5$ per unit area in distinct concentric annuli from the cluster center to the outermost region of the analyzed field of view. We have verified that the stellar density is constant for radial distance larger than $\sim$3 arcmin. This fact indicates that the number of cluster stars, in the reference field is negligible. 
The trichromatic image of the analyzed field is shown in Figure~\ref{fig:footprint}.
 
 The entire dataset is part of GO-14204 (PI A.\,P.\,Milone) which is a program specifically devoted to the study of multiple stellar populations in the young LMC clusters NGC\,1866 and NGC\,1755 (see Milone et al.\,2016b). All the images have been reduced and analyzed by using the method and software programs that have been mostly developed by Jay Anderson and are widely described in previous papers from this series. 

Briefly, we have first corrected the images for the effect of poor Charge Transfer Efficiency as in Anderson \& Bedin (2010) and then we have derived the stellar photometry and astrometry by using the software described in detail by Anderson et al.\,(2008) and adapted to UVIS/WFC3 images. Specifically, we have measured bright and faint stars by using a set of spatially-variable empirical point-spread functions (PSFs, see Anderson et al.\,2006 for details) but by adopting two different approaches. 
Fluxes of bright stars have been measured in each image independently, and the results combined later, while all the pixels of each very faint star in all the images have been fitted simultaneously. 
Stellar positions have been corrected for geometrical distortion by using the solution provided by Bellini, Anderson \& Bedin (2011) and photometry has been calibrated into the Vega-mag systems as in Bedin et al.\,(2005) and by adopting the zero points provided by the STScI web page for WFC3/UVIS\footnote{http://www.stsci.edu/hst/wfc3/phot\_{zp}\_{lbn}}.
  
The sample of stars used in our study of NGC\,1866 has been selected following Milone et al.\,(2009) and includes only relatively-isolated sources, that have been properly fitted by the PSF, and have small rms errors in position.  
 Finally, the photometry of stars with radial distances smaller than 3 arcmin from the cluster center have been corrected for differential reddening by following the recipe in Milone et al.\,(2012a) and adopting the values of $A_{\rm F336W}$ and $A_{\rm F814W}$ derived in Milone et al.\,(2016b).

In addition, we have used artificial stars (ASs) to determine the completeness level of our sample, to estimate internal photometric errors and to derive synthetic CMD. The AS tests have been performed following the procedure of Anderson et al.\,(2008), while the completeness has been determined as a function of both stellar position and magnitude by following the recipes of Milone et al.\,(2009).

\section{The color-magnitude diagram of NGC\,1866}
\label{sec:cmd}
\begin{centering} 
\begin{figure*} 
 \includegraphics[width=12.5cm]{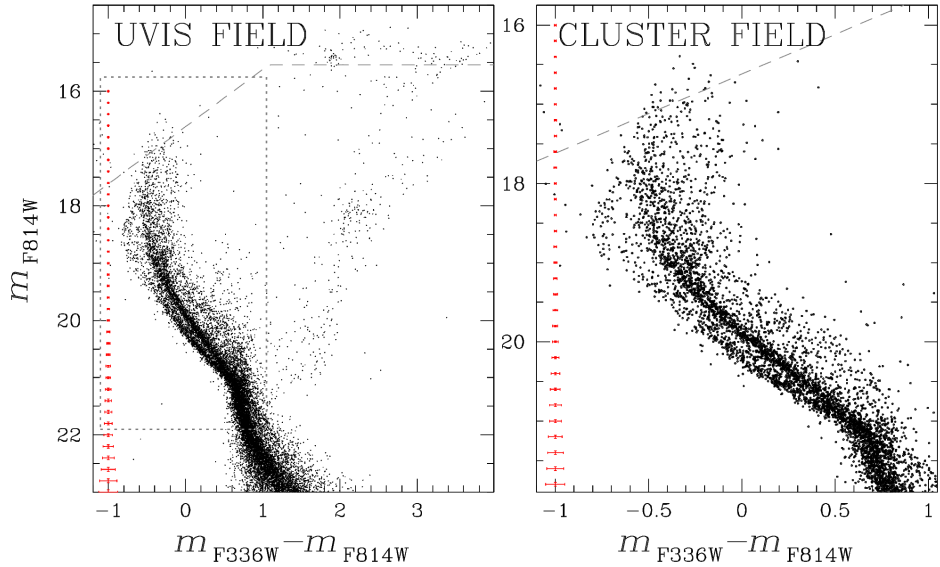} 
 \caption{\textit{Left panel:} $m_{\rm F814W}$ vs.\,$m_{\rm F336W}-m_{\rm F814W}$ CMD for all the stars in the WFC3/UVIS field of view. The photometry of stars above the dashed line has been obtained from saturated images in at least one filter.  
   \textit{Right panel:} Zoom in around the upper MS for stars with radial distance smaller than 41 arcsec from the center of NGC\,1866. The corresponding region of the CMD is marked by a dashed box in the left-panel plot. The error bars in red are shown on the left of each panel.} 
 \label{fig:cmd} 
\end{figure*} 
\end{centering} 
The $m_{\rm F814W}$ vs.\,$m_{\rm F336W}-m_{\rm F814W}$ CMD of all the stars in the WFC3/UVIS field of view is plotted in the left panel of Figure~\ref{fig:cmd} while in the right panel we show a zoom around the upper part of the MS of stars in the cluster field.
A visual inspection of these CMDs immediately reveals that the MSTO is broadened in color and magnitude in close analogy with what has been observed in NGC\,1856 and in the majority of the intermediate-age MC clusters (e.g.\,Mackey et al.\,2008; Milone et al.\,2009; Goudfrooij et al.\,2014). Furthermore, the upper MS is clearly split, and the two MSs merge together around $m_{\rm F814W}=21.0$. Noticeably, the red MS hosts the majority of MS stars, similarly to what we have observed in NGC\,1844, NGC\,1856, and NGC\,1755.

\begin{centering} 
\begin{figure*} 
  \includegraphics[width=12.5cm]{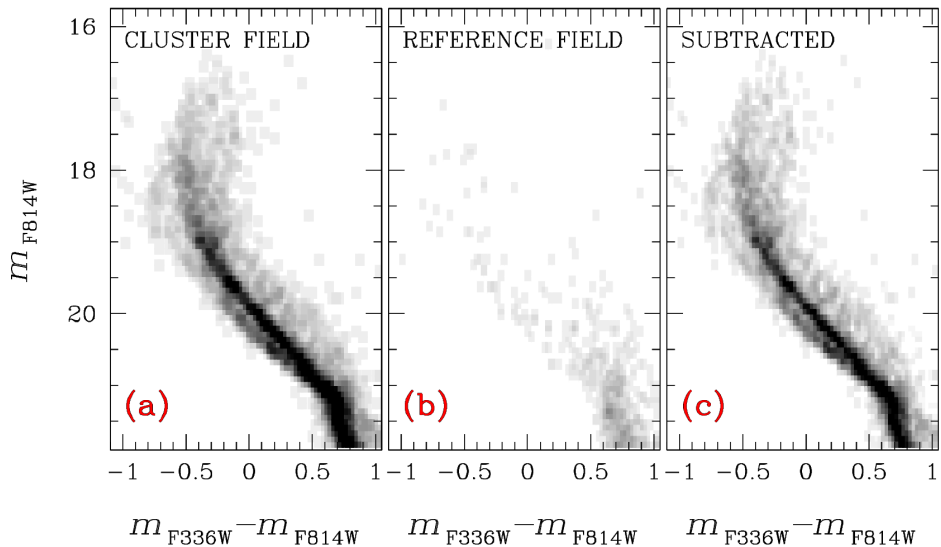}
 \caption{$m_{\rm F814W}$ vs.\,$m_{\rm F336W}-m_{\rm F814W}$ Hess diagram of stars in the cluster field (panel a) and in the reference field (panel b). The Hess diagram of the cluster CMD after field-stars have been subtracted is plotted in panel c.}
 \label{fig:hess} 
\end{figure*} 
\end{centering}

In the following we demonstrate that the split MS and the eMSTO are intrinsic features of NGC\,1866. 
We started by comparing the Hess diagrams of stars in the cluster field and in the reference field. These diagrams are plotted in the panels (a) and (b) of Figure~\ref{fig:hess}, respectively. The adopted level of gray used in this figure is proportional to the number of stars, corrected for completeness  and normalized to an area of one square arcsec, in each interval of color and magnitude. The panel (c) of Figure~\ref{fig:hess} shows the Hess diagram obtained by subtracting the star counts of the panel-(b) diagram from those of the panel-(a) one. 
 The fact that both the eMSTO and the split MS are present in the subtracted Hess diagram demonstrates that these features are real.

To further investigate the effect of field-star contamination on the cluster CMD we have compared in the panels (a) and (b) of Figure~\ref{fig:sub} the CMDs of stars in the cluster field and in the reference field. 
 In order to statistically subtract the stars of the reference-field CMD from the cluster-field CMD we have adapted to NGC\,1866 the same procedure used by Milone et al.\,(2009, 2015, 2016b). Specifically, we have determined for each star (i) in the reference field a distance in the $m_{\rm F814W}$ vs.\,$m_{\rm F336W}-m_{\rm F814W}$ CMD\\
{ \scriptsize $d_{\rm i}=\sqrt{k((m_{\rm F336W, cf}-m_{\rm F814W, cf})-(m^{\rm i}_{\rm F336W, rf}-m^{\rm i}_{\rm F814W, rf}))^{2}+(m_{\rm F814W, cf}-m^{\rm i}_{\rm F814W, rf})^{2}}   $}\\
 where
 $m_{\rm F336W (F814W), cf}$ and $m_{\rm F336W (F814W), rf}$ are the F336W (F814W) magnitudes in the cluster- and in the reference-field, respectively.
 The adopted constant $k=4.1$ accounts for the fact that the color of a star is better constrained than its magnitude (Gallart et al.\,2003) and has been determined as in Marino et al.\,(2014, see their Section~3.1).
 The stars in the cluster-field CMD  with the smallest distance to each star of the reference field have been considered as candidate to be subtracted. We have subtracted all the candidates with $r_{\rm i}<f c^{\rm i}_{\rm rf}/c^{\rm i}_{\rm cf}$ where $r_{\rm i}$ is a random number between 0 and 1,  $f$ is ratio between the area of the cluster field and of the reference field, and $c^{\rm i}_{\rm rf}$ and $c^{\rm i}_{\rm cf}$ are the completeness of the star (i) in the reference field and the completeness of the closest star in the cluster field, respectively.

 The decontaminated CMD is shown in panel (c) of Figure~\ref{fig:sub} and confirms that both the eMSTO and the split MS are intrinsic features of the cluster CMD. For completeness we show the CMD of the subtracted stars in the panel (d)  of Figure~\ref{fig:sub}.

\begin{centering} 
\begin{figure*} 
  \includegraphics[width=12.5cm]{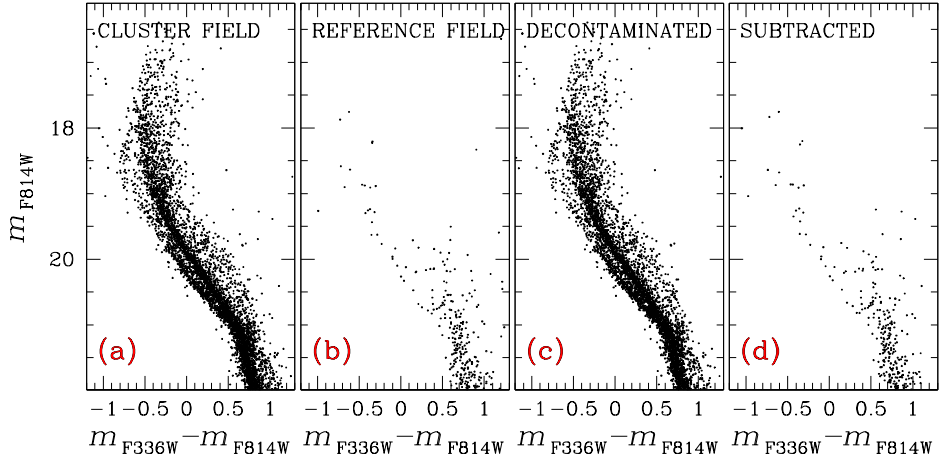}
  \label{fig:sub}
 \caption{Panel (a) reproduces the $m_{\rm F814W}$ vs.\,$m_{\rm F336W}-m_{\rm F814W}$ CMD of stars in the cluster field shown in the right panel of Fig.~\ref{fig:cmd}. The CMD of stars in the reference field is plotted in panel (b), while panel (c) shows the decontaminated CMD obtained by statistically subtracting the stars of the reference field from the cluster-field CMD. The CMD of the subtracted stars is plotted in panel (d).}
 \label{fig:sub} 
\end{figure*} 
\end{centering} 

\section{The double MS}
\label{sec:ms}
Having demonstrated that the split MS of NGC\,1866 is real, we estimate in the following the fraction of stars in each sequence and the binary fraction.  The population ratio as a function of the stellar luminosity and the radial distributions of red-MS and blue-MS stars are derived in Section~\ref{sub:rd} and ~\ref{sub:lf}, respectively.

\begin{centering} 
\begin{figure*} 
 \includegraphics[width=13cm]{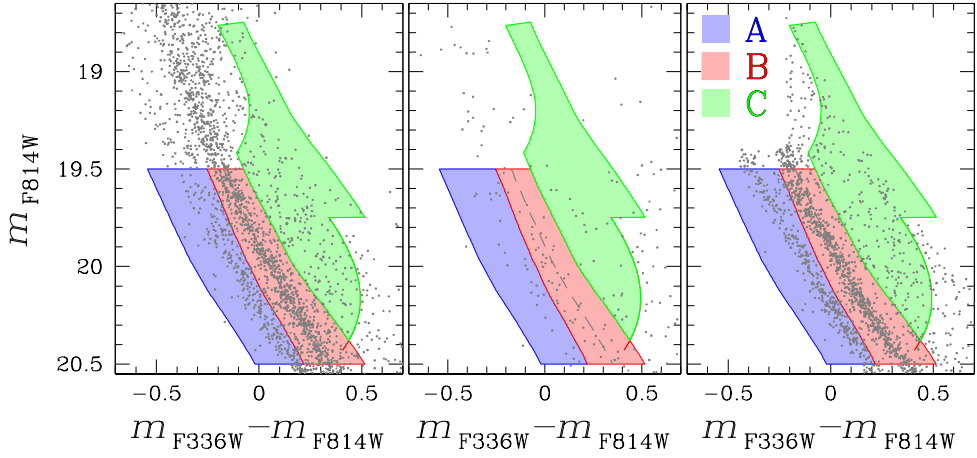} 
 \caption{Zoom in of the $m_{\rm F814W}$ vs.\,$m_{\rm F336W}-m_{\rm F814W}$ CMD around the region where the MS split is most prominent. Left and middle panel show the observed CMD for stars in the cluster and the reference field, respectively. The simulated CMD that best reproduces the observed ones is plotted in the right-panel. The shaded areas are the CMD regions A, B, C used to determine the fraction of red-MS and blue-MS stars and the binary fraction. See text for details.}
 \label{fig:METpratio} 
\end{figure*} 
\end{centering} 

In order to infer the fraction of red-MS stars, blue-MS stars, and the fraction of binaries in the cluster field we have adapted to NGC\,1866 the method described by Milone et al.\,(2012b, MPB12 hereafter). To do this we have defined three regions in the CMD, namely A, B, and C which are represented by the blue, red, and green shadow area, respectively, in Figure~\ref{fig:METpratio}.
 These three regions have been derived as follows.
 Regions A and B are mostly populated by blue-MS stars and red-MS stars with $19.5<m_{\rm F814W}<20.5$, respectively. Note that in the adopted magnitude interval the two MSs are clearly split and binaries with mass ratio q$>0.5$ are well separated from single MS stars.
 The blue boundary of region A has been drawn arbitrarily with the criteria of including the majority of blue-MS stars. Its red boundary has been determined by shifting each point of the red-MS fiducial line 
 by 2$\sigma_{\rm color}$ towards blue colors, where $\sigma_{\rm color}$ is the uncertainty in the $F336W-F814W$ color determination. The red-MS fiducial line has been derived by following the procedure described in the Section~3.3 of Milone et al.\,(2016b). 
Briefly,  we have defined a series of $F814W$ magnitude bins of width $\nu=0.1$ mag in the interval with $18.9<m_{\rm F814W}<20.9$. These bins have been determined over a sample of $N$ points separated by steps, $s=\nu/3$, of fixed magnitude (see Silverman 1986 for details). 
Then we have selected a sample of bona-fide red-MS stars and calculated the median color and their mean magnitudes in each magnitude interval. The red-MS fiducial has been obtained by interpolating these median color and mean magnitudes by means of a cubic spline.
 
 The region C defined in  Figure~\ref{fig:METpratio} mostly includes binary systems with large mass ratio and has been derived as in MPB12. Its blue boundary is the sequence of binary systems with mass ratio q=0.5 formed by two red-MS stars. The red boundary of region C corresponds to the sequence of equal-mass red-MS binaries shifted to the red by four times the error in color. The faint and the bright boundaries correspond to the locus populated by a binary system formed by two red-MS stars where the primary component has luminosity $m_{\rm F814W}=20.5$ and $m_{\rm F814W}=19.5$. Region B is placed between regions A and C.

 We have determined the number of stars, corrected for completeness, in the regions A, B, and C of the cluster-field CMD ($N^{\rm CL-F}_{\rm A, B, C}$) and of the reference-field CMD ($N^{\rm REF-F}_{\rm A, B, C}$). We have estimated the number of cluster stars in each region as $N^{\rm CL}_{\rm A, B, C}=N^{\rm CL-F}_{\rm A, B, C}- f N^{\rm REF-F}_{\rm A, B, C}$, where $f$ is ratio between the area of the cluster field and of the reference field, respectively.

 We have generated a large number of CMDs by using ASs and compared them with the observed CMD. Each simulated CMD hosts the same number of region-B stars ($N^{\rm CL}_{\rm B}$) as the observed CMD but includes different fractions of blue-MS stars, red-MS stars, and binaries ($f_{\rm bMS}$, $f_{\rm rMS}$, and $f_{\rm bin}$). 
 Specifically, the grid of simulated CMDs have $f_{\rm bMS}$ and $f_{\rm bin}$ ranging from 0.01 to 1.00 in steps of 0.01.
 Binaries have been added by assuming a constant mass-ratio distribution, in close analogy with what has been observed in Galactic GCs (Milone et al.\,2012a; 2016c). Moreover we have assumed that both the red and the blue MS have the same binary fraction and that both components of each binary system belong to the same sequence.

 To obtain the best match between the simulated and the observed CMDs, we imposed that the simulated CMDs have the same number of stars in the regions A, B, and C as the observed ones. This condition is satisfied when the blue MS hosts 30$\pm$2\% of the total number of analyzed MS stars and for $f_{\rm bin}=0.25 \pm 0.02$.
 For completeness, we have extended the analysis to the entire region with radial distance from the cluster center smaller than 3.0 arcmin and find slightly-higher values for both the fraction of blue MS stars ($f_{\rm bMS}=0.35 \pm 0.02$) and the binary fraction ($f_{\rm bin}=0.28 \pm 0.02$).

 \subsection{The population ratio as a function of the stellar luminosity}
 \label{sub:lf}
 
\begin{centering} 
\begin{figure*} 
 \includegraphics[width=12.5cm]{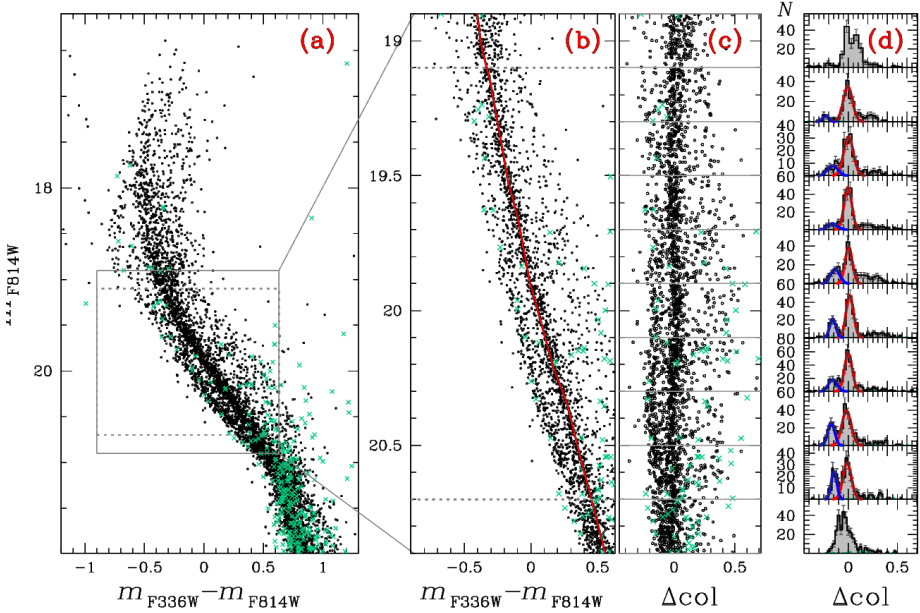} 
 \caption{This figure illustrates the procedure used to study the color distribution of MS stars in NGC\,1866. Panel (a) shows the CMD of stars in the cluster field (black points) and in the reference field (green crosses) while panel (b) is a zoom of the analyzed MS region. The red line is the fiducial line of the red MS and has been used to derive the verticalized $m_{\rm F336W}$ vs.\,$\Delta$($m_{\rm F336W}-m_{\rm F814W}$) CMD plotted in the panel (c). Panel (d) shows the histogram distribution of $\Delta$($m_{\rm F336W}-m_{\rm F814W}$) for stars in the ten F814W magnitude intervals indicated by continuous lines in the panel (c). The histogram distribution of stars with $19.1<m_{\rm F814W}<20.7$ is clearly bimodal. The best fit least-squares bi-Gaussian function is overimposed to the histograms and the two Gaussian components are colored blue and red (see text for details).  }
 \label{fig:pratio} 
\end{figure*} 
\end{centering}

 To investigate the multiple stellar populations along the double MS of NGC\,1866, we started to analyze in Figure~\ref{fig:pratio} the $m_{\rm F814W}$ vs.\,$m_{\rm F336W}-m_{\rm F814W}$ CMD of cluster-field (black points) and reference-field (aqua) stars with $18.9<m_{\rm F814W}<20.9$.
  Only field stars with $r_{\rm i}<f c^{\rm i}_{\rm rf}/c^{\rm i}_{\rm cf}$ have been used in the following analysis, in close analogy with what we have done in Section~\ref{sec:cmd}.
 We plotted in the panel (b) of Figure~\ref{fig:pratio}, a zoom of the CMD for the stars located within the gray box in the CMD shown in the panel (a).  The red line is the red-MS fiducial derived as in Section~\ref{sec:ms}.
 Panel (c) of Figure~\ref{fig:pratio} shows the verticalized $m_{\rm F814W}$ vs.\,$\Delta$($m_{\rm F336W}-m_{\rm F814W}$) CMD. The latter quantity has been obtained by subtracting from the $m_{\rm F336W}-m_{\rm F814W}$ color of each star the color of the fiducial at the corresponding $F814W$ magnitude. 

 The $\Delta$($m_{\rm F336W}-m_{\rm F814W}$) histogram distribution of cluster-field stars in ten magnitude bins is provided in the panels (d) and confirms the visual impression that the MS is bimodal in the magnitude interval $19.1<m_{\rm F814W}<20.7$. We have used a bi-Gaussian function to fit by means of least squares the observed histogram distribution, after the small contribution from reference-field stars has been subtracted. The two Gaussian components that best fit the histograms have been represented with blue and red continuous lines. From the area under the Gaussians we have inferred the fraction of stars in each interval of magnitude. 
 
 The two MS merge together around $m_{\rm F814W} \sim 20.7$ and there is no evidence for a split MS at fainter luminosities. The $\Delta$($m_{\rm F336W}-m_{\rm F814W}$) spread significantly increases for $m_{\rm F814W} \lesssim 19.1$, mainly because the analyzed sample includes stars in the faintest part of the eMSTO. A small number of blue-MS stars is still present in this luminosity interval.

 To investigate the properties of the split MS at different luminosities by means of a different method, we have divided the MS region within $19.5<m_{\rm F814W}<20.5$ into five intervals of 0.2 mag and we have determined the fraction of blue-MS stars and the fraction of binaries with respect to the total number of MS stars in each of them by using the procedure from MPB12 as described in Section~\ref{sec:ms}. In this case we have excluded from the analysis the upper MS, where it is not possible to distinguish binaries with q$>$0.5 from single MS stars. We have also excluded the faintest MS part due to the poor separation between the blue and the red MS.

\begin{centering} 
\begin{figure} 
 \includegraphics[width=8.5cm]{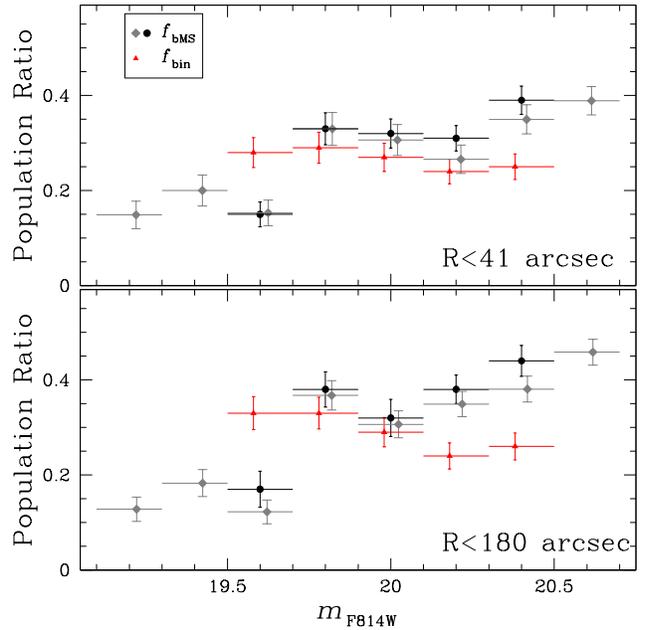} 
 \caption{Fraction of blue-MS stars (black dots and grey diamonds) and fraction of binaries (red triangles) with respect to the total number of MS stars in different F814W magnitude intervals.
   Black dots and grey diamonds indicate the results from the MPB12 method and by using the Gaussian-fit method, respectively.
   The upper panel refer to the cluster field, while the population ratios derived in the region with radial distance smaller than 180 arcsec are plotted in the lower panel. For clarity the grey and red points have been shifted by $\pm$0.05 mag with respect to the average magnitude of the stars in the corresponding bin.}
 \label{fig:LF} 
\end{figure} 
\end{centering} 
 The results are illustrated in the upper panel of Figure~\ref{fig:LF}, where we plot the fraction of blue-MS stars with respect to the total number of MS stars in the cluster field as a function of the F814W magnitude. The grey diamonds indicate the results from the method based on bi-Gaussian fitting while the black dots are obtained from the procedure by MPB12. Red triangles show the derived fraction of binaries as a function of $m_{\rm F814W}$.

 The two methods point to similar conclusions. We find that the blue-MS hosts $\sim$15\% of the total number of MS stars in the brightest analyzed magnitude bin and that the fraction of blue-MS stars rises up to $\sim$33\% for $m_{\rm F814W}>19.7$. We did not find any evidence for a significant variation of the binary fraction with the F814W magnitude in the analyzed luminosity interval.

 For completeness, we have extended the analysis to the entire region with radial distance smaller than 3.0 arcmin from the cluster center. The resulting values of $f_{\rm bMS}$ and $f_{\rm bin}$ are shown in the bottom panel of Figure~\ref{fig:LF} and confirm the conclusions obtained from the cluster field.  

 \subsection{The radial distribution of the two MSs}
 \label{sub:rd}

\begin{centering} 
\begin{figure*} 
 \includegraphics[width=13.5cm]{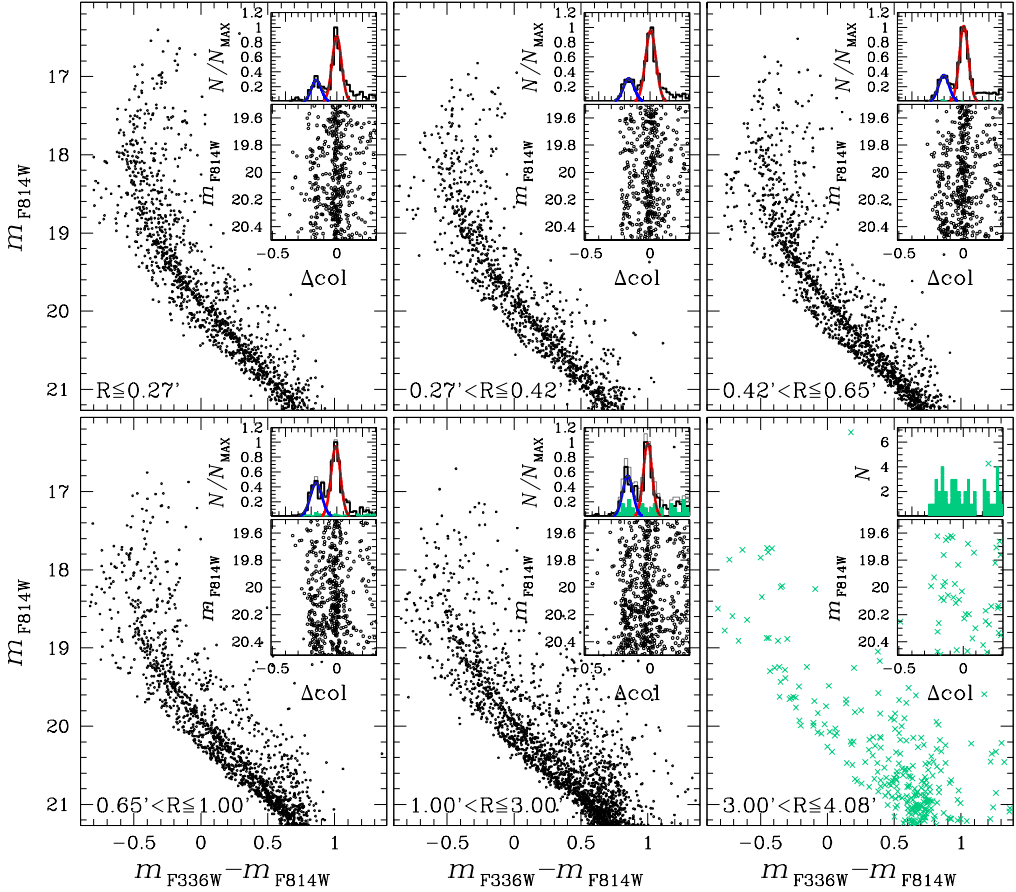} 
 \caption{$m_{\rm F814W}$ vs.\,$m_{\rm F336W}-m_{\rm F814W}$ CMDs of stars with different radial distance from the center of NGC\,1866. The verticalized $m_{\rm F814W}$ vs.\,$\Delta$~col diagram for stars with $19.5<m_{\rm F814W}<20.5$ is plotted in the lower inset of each panel. 
   For the panels of stars within three arcmin from the cluster center we show the normalized $\Delta$~col histogram distribution for all the stars in the inset  (gray line), the normalized histogram for field stars (aqua line), and the normalized histogram for cluster members (black line). The shaded aqua histograms correspond to reference-field stars only.
   The red and blue lines overimposed on the black histogram are the two components of the best-fit bi-Gaussian function.
Reference-field stars are represented with aqua crosses in the bottom-right panel where we also show the $\Delta$~col histogram distribution for all the stars in the inset.}
 \label{fig:RDcmd} 
\end{figure*} 
\end{centering} 

 In order to investigate how the CMD morphology changes as a function of cluster-centric radius we have employed two methods, in close analogy with what we have done in the study of multiple populations along the MS. The first method, is based on bi-Gaussian fit of the color distribution of the MS and is illustrated in Figure~\ref{fig:RDcmd} where we plot the CMDs of stars in six annuli with different distances from the cluster center. The inner and the outer radius of each annulus have been chosen in such a way that each CMD includes the same number of stars in the region B. The inset of each panel of Figure~\ref{fig:RDcmd} shows the verticalized $m_{\rm F814W}$ vs.\,$\Delta$~col diagram for stars in the magnitude interval with $19.5<m_{\rm F814W}<20.5$ where the MS split is more evident and where the two MSs run almost parallel.

In the insets of the five CMDs of stars with R$<$3.0 arcmin we also plot the corresponding normalized $\Delta$~col histogram distribution (grey line). Moreover we show the histogram of the field stars expected in each annulus (aqua histograms) and the histogram of cluster members (black line) which has been  obtained by subtracting the aqua histogram from the grey ones. We have used a bi-Gaussian function to match the black histograms by means of least squares and we have colored the two components of the best-fitting bi-Gaussians red and blue. All the histograms in each panel have been normalized to the maximum value of the histogram of cluster stars. 

The lower-right panel shows reference-field stars with radial distance from the cluster center larger than 3.0 arcmin. The corresponding histogram of the $\Delta$ col distribution is shown in the inset.

\begin{centering} 
\begin{figure*} 
 \includegraphics[width=8.5cm]{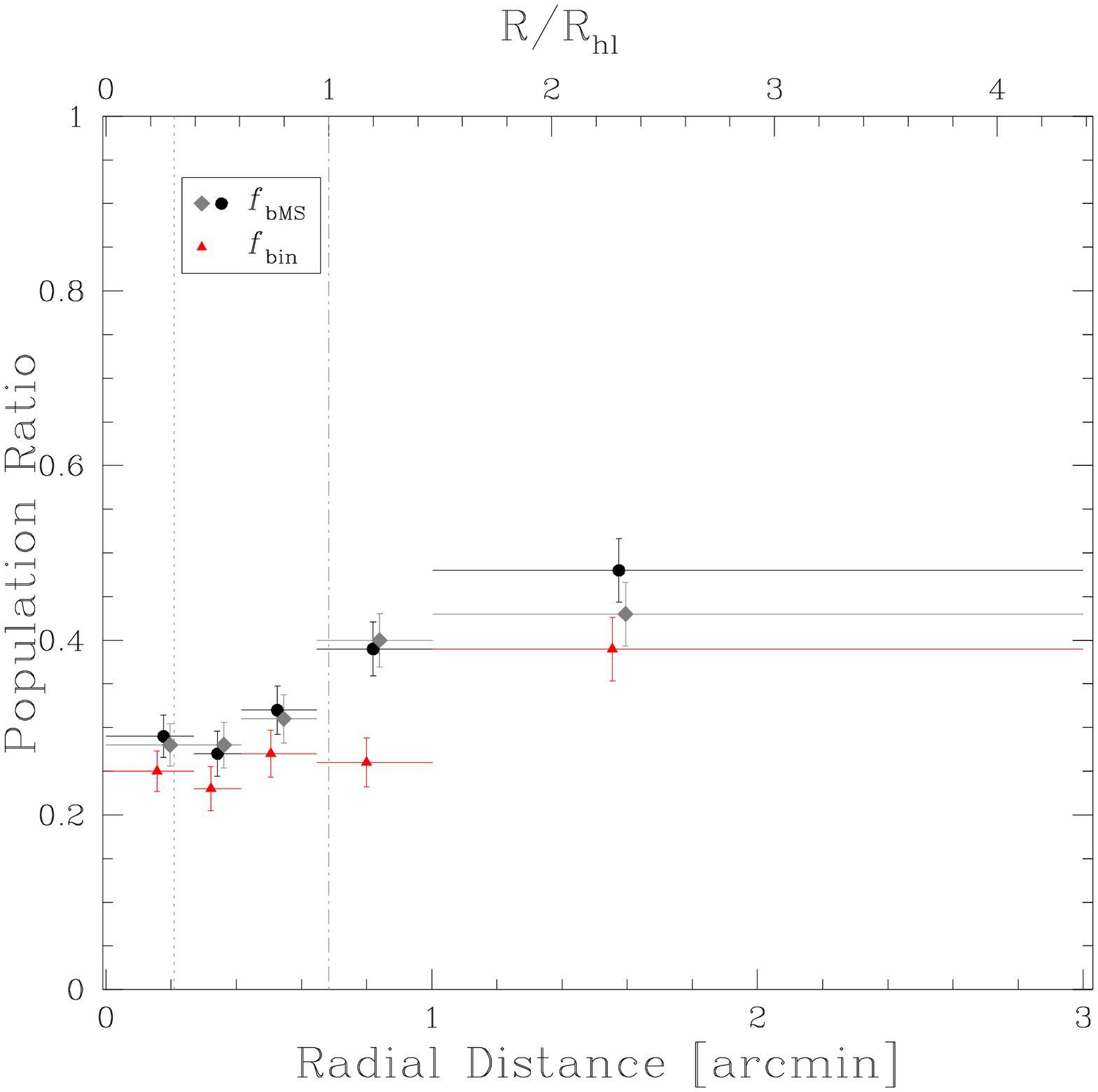} 
 \includegraphics[width=8.5cm]{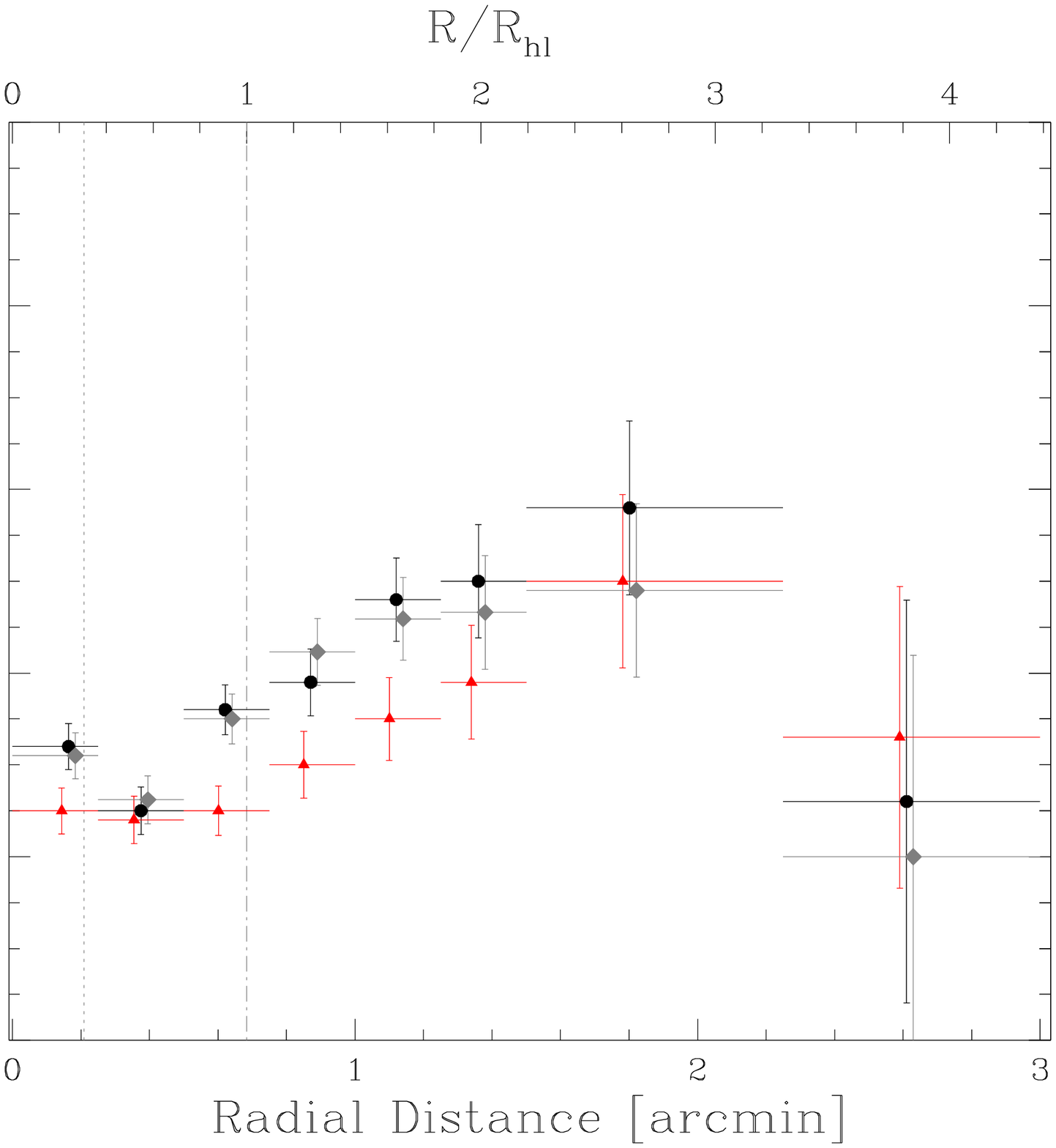} 
 \caption{Fraction of blue-MS stars with respect to the total number of analyzed MS stars as a function of the radial distance from the cluster center (in arcmins and in units of projected half-light radius, $R_{\rm hl}$). The black and grey points with error bars indicate the results obtained from the MPB12 method and the bi-Gaussian fit method, respectively, as described in the text. Red points show the fraction of binaries in different radial intervals. For clarity, the grey and red points have been shifted by $\pm$0.03 arcmin with respect to the average radius of stars in each radial bin. The horizontal bars mark the radial extension of each bin. The dotted line and  the dashed-dotted vertical line mark the projected core and half-light radius, respectively (Mc Laughlin \& Van der Marel\,2005).  In the left- and right-panel plot we have used different radial bins. See text for details.}
 \label{fig:RDngc1866} 
\end{figure*} 
\end{centering}

 Figure~\ref{fig:RDcmd} suggests that the fraction of blue-MS stars with respect to the total number of MS stars significantly increases when we move from the cluster center outwards. In particular, from the area under the Gaussians, we find that in the central regions $\sim$30\% of the total number of MS stars belong to the blue MS, while, in the outermost analyzed region, the fraction of blue-MS stars rises to $\sim$45\%. The grey diamonds plotted in  Figure~\ref{fig:RDngc1866} show the resulting fraction of blue MS stars ($f_{\rm bMS}$) as a function of the average radial distance from the cluster center of all the stars in the bin.  

 To further investigate the radial distribution of the stellar populations in NGC\,1866, we have used the recipe by MPB12 described in Section~\ref{sec:ms} to determine the fraction of red-MS and blue-MS stars and the fraction of binaries with respect to the total number of MS stars in each annulus defined above.   
 Results are illustrated in  the left panel of Figure~\ref{fig:RDngc1866} and confirm that the red MS is more centrally concentrated than blue-MS stars. As shown in Figure~\ref{fig:RDngc1866}, the binary fraction has a flat distribution for cluster-centric distance smaller than R$\sim 1$ arcmin where their fraction is around 25\%. In the cluster outskirts the binary fraction rises up to $f_{\rm bin}=0.38 \pm 0.04$ but this result is significant at the 2-$\sigma$ level only.

 Moreover, we have investigated the radial distribution of the stellar populatations by using  radial bins of fixed size. Specifically, we have chosen 0.25-arcmin radial bins for stars within 1.5 arcmin from the cluster center and 0.50-arcmin bins for R$>\geq 1.5$ arcmin. The large size of the two outermost bins is due to the small number of stars in the external cluster region. Results are illustrated in the right panel of Figure~\ref{fig:RDngc1866} and confirm the previous finding that red-MS stars are more centrally-concentrated than blue-MS stars. This figure corroborates the idea that the binary fraction increases towards the cluster outskirts and makes it tempting to speculate that a large fraction of binaries are thus associated with the blue MS. A possible exception to these trends is provided by stars in the outer-most radial bin with $2.25<R\leq 3.0$ arcmin but the large error bars prevent us from any firm conclusion.

\section{Theoretical interpretation}
\label{sec:teo}
The eMSTO and the split MS of young and intermediate-age star clusters have been interpreted both in terms of stellar populations with different rotation rates (e.g.\,Bastian \& De Mink\,2009; Niederhofer et al.\,2014; D'Antona et al.\,2015; Milone et al.\,2016a) or with different ages (e.g.\,Mackey \& Broby Nielsen\,2007; Milone et al.\,2009; Goudfrooij et al.\,2014). In addition, Milone et al.\,(2015) have shown that the eMSTO and the double MS of NGC\,1856 are consistent with stellar populations with different metallicity. 

In NGC\,1866, we can immediately exclude that internal metallicity variations are responsible for the eMSTO and the split MS. Indeed high-resolution spectroscopy of 14 cluster members has revealed neither significant iron spread nor evidence for star-to-star variation of a variety of light elements including sodium, oxygen, and magnesium (Mucciarelli et al.\,2011). Similarly, the analysis of eight stars in the eMSTO cluster NGC\,1806 shows no evidence for metallicity variations (Mucciarelli et al.\,2014, Mackey et al.\,in preparation). These results demonstrate that the eMSTO and the split MS of NGC\,1866, and the eMSTO of NGC\,1806 are not due to stellar populations with different Z, and this suggests that similar features observed in the CMDs of other clusters are unlikely due to stars with different metallicity.

In the following subsection we compare the observed CMD of NGC\,1866 with isochrones from the Geneva database\footnote{http://obswww.unige.ch/Recherche/evoldb/index} (Mowlavi et al.\,2012; Ekstr{\"o}m et al.\,2013; Georgy et al.\,2014) that correspond to stellar populations with different ages, while in Section~\ref{subsec:rot} we investigate the possibility that the eMSTO and the double MS of NGC\,1866 are due to different rotation rates. In the latter subsection we will also compare the observations with isochrones and synthetic CMDs with both different age and rotation rates.

The comparison between the isochrones and the observed CMD of stars in the cluster field are shown in Figures~\ref{fig:iso}.
 The adopted values of reddening and distance modulus are quoted in the inset of each panel. The reddening values have been transformed into absorption in the $F336W$ and $F814W$ bands by using the relations between E($B-V$), A$_{\rm F336W}$, and A$_{\rm F814W}$ derived by Milone et al.\,(2016a).
 
 \subsection{Age variation}
 \label{subsec:age}
 The comparison between the observed CMD and non-rotating isochrones with the same metallicity but different ages is shown in the left panel of Figure~\ref{fig:iso}. First we have determined that the blue MS and the brighter eMSTO is well reproduced by a 140-Myr-old population with metallicity Z=0.006, assuming a distance modulus (m$-$M)${_0}$=18.31 and reddening $E(B-V)$=0.11. Then we have searched for an isochrone with the same value of $Z$ that properly fits both the red MS and lower part of the eMSTO. 
 We find that a 220-Myr-old isochrone matches well the lower eMSTO but provides a very poor fit of the red MS. We conclude that age variation alone can not be the responsible for the eMSTO and the split MS of NGC\,1866.   
 
 \subsection{Rotation}
 \label{subsec:rot}
\begin{centering} 
\begin{figure*} 
 \includegraphics[width=8.45cm]{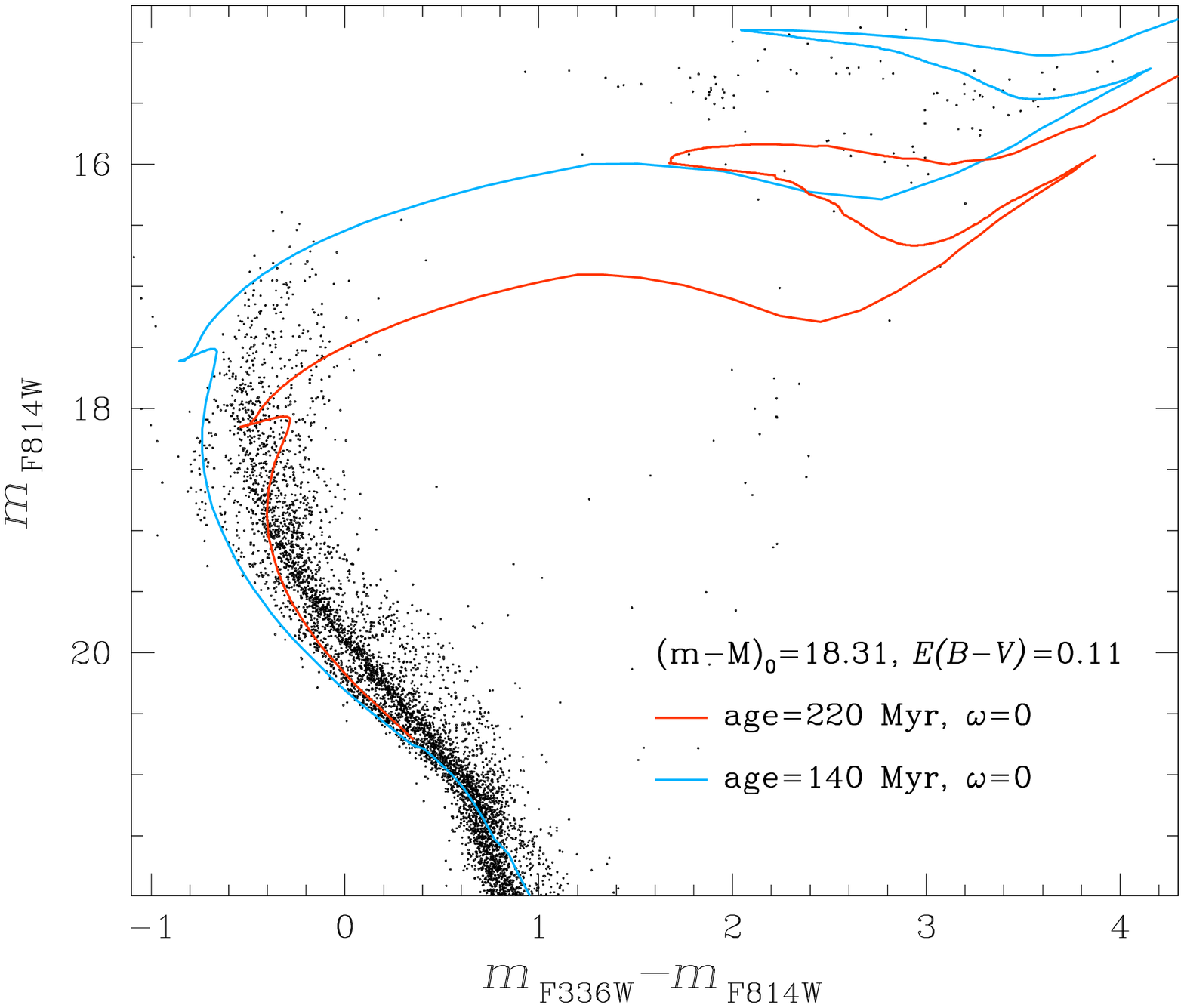} 
 \includegraphics[width=7.5cm]{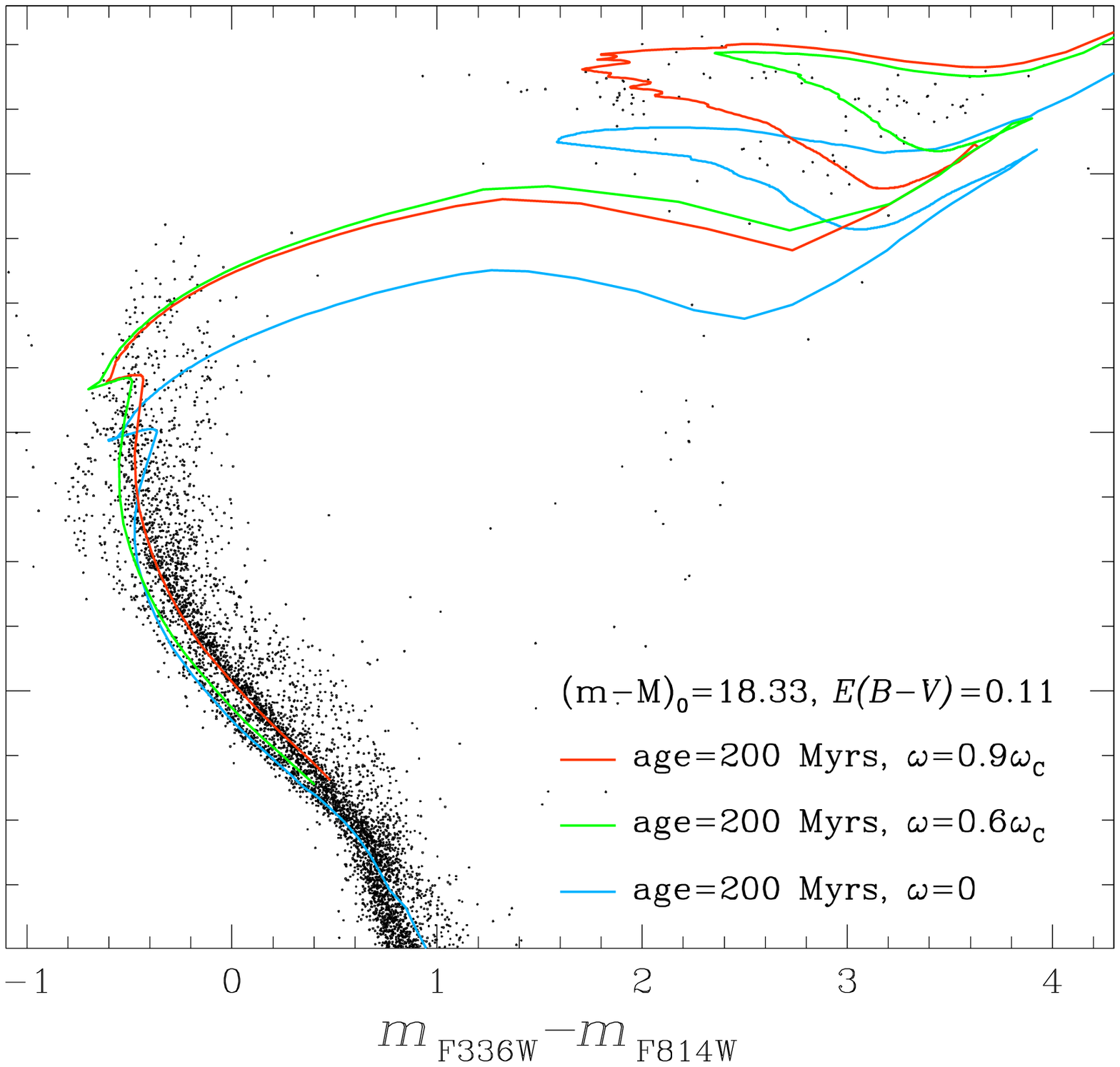} 
 \caption{Comparison between the observed CMD of stars in the cluster field and isochrones from the Geneva database. In the left panel we have plotted two non-rotating isochrones with different ages while in the right panel we show three coeval isochrones with no rotation and with rotations of $\omega=0.6 \omega_{\rm c}$ and $\omega=0.9 \omega_{\rm c}$. } 
 \label{fig:iso} 
\end{figure*} 
\end{centering}

 In the right panel of Figure~\ref{fig:iso} we investigate the possibility that rotation is the only factor responsible for the double MS and the eMSTO of NGC\,1866 by using isochrones with the same age of 200 Myr and different rotation rates of $\omega=0$ (blue line), $\omega=0.6 \omega_{\rm c}$ (green line, where $\omega_{\rm c}$ is the critical rotation value), and $\omega=0.9 \omega_{\rm c}$ (red line). We find that the red MS is well fitted by the fast-rotating stellar population  while the blue MS is reproduced by a non-rotating isochrone. Nevertheless, rotation alone does not reproduce the upper part of the blue MS with $m_{\rm F814W} \lesssim 19$.


\begin{centering} 
\begin{figure*} 
 \includegraphics[width=13.0cm]{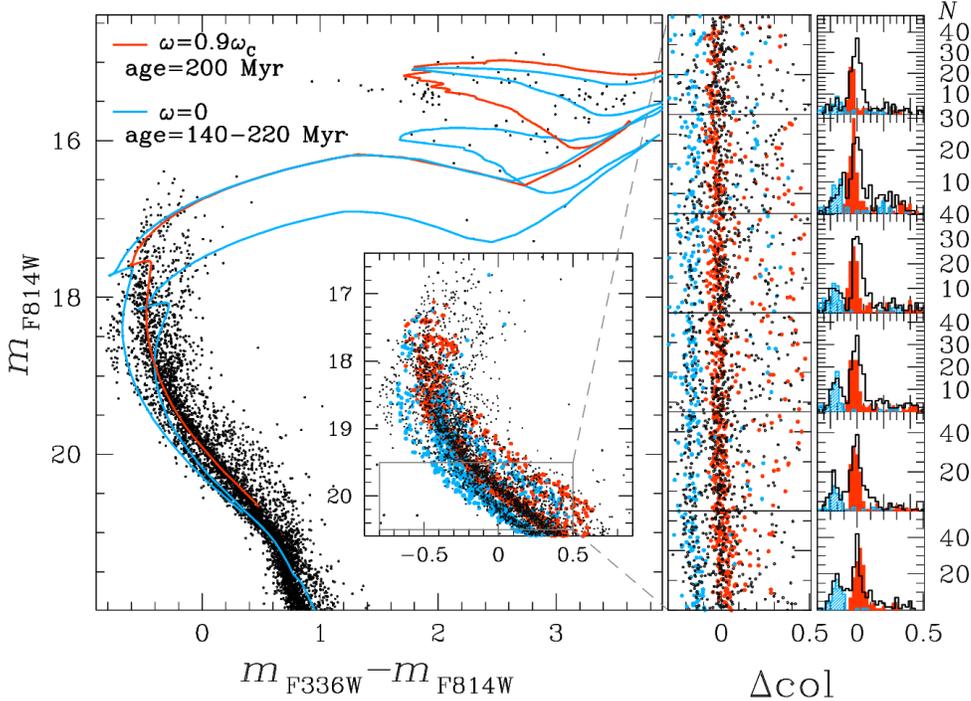}
 \caption{The large panel shows three isochrones from the Geneva database with different ages and rotation rates, overimposed on the CMD of stars in the cluster field. The red line corresponds to the isochrone with age of 200 Myr and has a rotational velocity close to the breakout value $\omega=0.9 \omega_{\rm c}$. The two non-rotating isochrones are colored cyan and have ages of 140 and 220 Myrs. The inset shows a zoom around the upper MS of NGC\,1866, where we compare the simulated CMD derived from the rotating (red) and non-rotating (cyan) isochrones and the observed CMD (black). The corresponding verticalized $m_{\rm F814W}$ vs.\,$\Delta$~col is plotted in the middle panel, while in the right panels we compare the histogram $\Delta$~col distribution of the observed stars in  five magnitude intervals and the distribution of the simulated rotating and rotating stars. See text for details.} 
 \label{fig:Rotage} 
\end{figure*} 
\end{centering}

Finally, we investigate in Figure~\ref{fig:Rotage} the possibility that the observations are consistent with stellar populations with both different ages and different rotation rates.
 The large panel of this figure shows that a 200-Myr isochrone with rotation $\omega=0.9 \omega_{\rm c}$ (red line) provides a good fit at the red MS, while the blue MS is well reproduced by two non-rotating isochrones with age of 140 Myr and 220 Myr. %

 In the inset we compare the observed CMD of NGC\,1866 with a synthetic CMD that  includes stellar populations with both age variation and different rotation rates.  For that purpose we have retrieved from the Geneva database the isochrones plotted in the left panel. Specifically we have assumed that 70\% of total number of the simulated stars have rotation $\omega=0.9 \omega_{\rm c}$ and the remaining 30\% of stars do not rotate. Among them two-thirds have age of 220 Myr and the remaing one-third belong to a younger 140-Myr old population.

  The viewing angle adopted in the simulations follow a random distribution, and the adopted gravity-darkening model from Espinosa Lara \& Rieutord (2011) includes the limb-darkening effect (Clare 2000).  The synthetic data have been transformed into the observational plane by using the model atmospheres by Castelli \& Kurucz (2003) and the transmission curves of the $F336W$ and $F814W$ filters of UVIS/WFC3. The simulated CMD includes a fraction of binaries, $f_{\rm bin}=0.25$, which corresponds to the observed value. We have used blue and red colors to represent non-rotating and rotating stars, respectively, while the observed stars are colored in black.

  In the middle panel of Figure~\ref{fig:Rotage} we compare the observed and the simulated verticalized $m_{\rm F814W}$ vs.\,$\Delta$~col diagram for MS stars with $19.5<m_{\rm F814W}<20.5$. The corresponding $\Delta$~col histogram distributions of stars in five magnitude intervals are plotted in the right panels. These figures show that the adopted synthetic models reproduce well the double MS of NGC\,1866.

  An eMSTO is present in the synthetic CMD, but the fit with the observed CMD is still unsatisfactory. This issue has been previously noticed in both NGC\,1856 and NGC\,1755 and has been attributed to second-order parameters that affect the hydrogen-burning phase (D'Antona et al.\,2015). Indeed the stellar color and magnitude of stars in the synthetic CMD are strongly affected by the way the convective-core overshoot and the inclination angle are taken into account.

\section{Summary and Discussion}
\label{sec:discussion} 

We have used {\it HST\,} to derive high-precision photometry of the young LMC cluster NGC\,1866 in the $F336W$ and $F814W$ bands of WFC3/UVIS.
The resulting CMD reveals that this cluster has a double MS with the blue component hosting about one third of the total number of MS stars in the analyzed magnitude interval.
A bimodal MS has been recently observed in other LMC clusters younger than $\sim$400 Myr including NGC\,1844, NGC\,1856 and NGC\,1755 (Milone et al.\,2013, 2015, 2016a). The finding of a similar feature in NGC\,1866 corroborates the hypothesis that the split MS  is a common feature of young Magellanic Cloud star clusters.
 In addition, NGC\,1866 exhibits an eMSTO in close analogy with what is observed in most intermediate-age stars clusters of both Magellanic Clouds and in some young LMC clusters (Mackey \& Broby Nielsen 2007; Goudfrooij et al.\,2011, 2014; Milone et al.\,2009, 2015, 2016a; Bastian et al.\,2016).

 The relative numbers of blue- and red-MS stars change when moving from the cluster center to the external regions, with the red MS being more-centrally concentrated. While this is the first study on the radial distribution of the multiple MSs in a young cluster, the radial distribution of eMSTO stars in intermediate-age star clusters has been already determined for several clusters by Goudfrooij et al.\,(2011).
  These authors find that for several massive clusters the stars in the brightest half of the MSTO region are significantly more centrally concentrated than the stars in the faintest half. 

  In the cluster field of view the binary fraction with respect to the total number of MS stars of NGC\,1866 is $f_{\rm bin}=0.25 \pm 0.02$.
  The binary fraction is almost constant within $\sim$1 arcmin from the cluster center and seems to increase at larger radial distance, in analogy with what has been observed in the massive young LMC cluster NGC\,1805 by Li et al.\,(2013). 

  The comparison between stellar models and the observed CMD rules out the possibility that age variation is the only responsible for the split MS of NGC\,1866.
 In contrast, the split MS is well matched by stellar populations with distinct rotation rates.
 Specifically, the red MS is consistent with a $\sim$200-Myr old population of fast-rotating stars with $\omega=0.9 \omega_{\rm c}$ while the blue MS is reproduced by non-rotating stars.
  The adopted stellar models with different age and rotation rates roughly reproduce the eMSTO. As suggested by D'Antona et al.\,(2015), we speculate that the poor fit is mostly due to second-order parameters adopted in the models that affect the hydrogen-burning phase.
 
 Noticeably, it appears that rotation alone is not able to fully reproduce the observations of NGC\,1866. Indeed, while the majority of blue-MS stars are  possibly coeval,  or about 10\% older than the red MS, the upper part of the blue MS is reproduced by a $\sim 30$\% younger ($\sim$140-Myr old) stellar population including $\sim$15\% of the total number of MS stars.

  We are aware that the results presented in this work open many more questions than they solve. We regard the preliminary reproduction of the observational features of the CMD, proposed here and shown in Figure~\ref{fig:Rotage}, more as a provocative message than as a trustful interpretation.  Is it realistic to think that stars born in possibly three different epoch of star formations ($\sim$140, $\sim$200 and $\sim$220 Myr ago) are present in the cluster? The first consequence of this scenario would be that the first star formation event gives birth to a non-rotating population, the second one to a fast rotating population and the final burst, $\sim$60 Myr later, would include only non rotating stars.  This scheme seems too complex to be realistic, and we should reject it. Nevertheless, the multiple-age scheme may be telling us something on the evolution of these stars. 

  Also the increase of the blue MS fraction with increasing distance from the cluster center would argue against the younger age of the blue population.
 Indeed, if there is an analogy with the multiple populations in old GCs, we expect that second-generation stars are expected to be more centrally concentrated than the first populations (e.g.\,D'Ercole et al.\,2016 and references therein).

 Our interpretation of the split MS of NGC\,1866 are similar to those by D'Antona et al.\,(2015) who have shown that the split MS and the eMSTO of the $\sim$350-Myr old LMC cluster NGC\,1856 are consistent with two stellar populations with different rotation rates. In this framework, D'Antona and collaborators attributed the presence of a fraction of non-rotating stars to the braking mechanism of dynamical tides acting on the convective H-burning cores of B-A type stars due to the presence of a binary companions (Zahn 1977). In fact, observations show that binaries of these spectral types, with periods from 4 to 500 days are synchronized, so they are all slowly rotating (Abt \& Boonyarak 2004). If this phenomenon is the responsible for the slowly rotating fraction of stars, we could expect that there are more binaries with such adequate periods in the external parts of the cluster. 

  While it can be straightforward to interpret the populations in the cluster  NGC\,1856, about a factor two older, by means of two coeval populations, with 70\% of fast rotating stars, and $\sim$30\% of slowly rotating stars, (D'Antona et al.\,2015, see above), the case of NGC\,1866 looks much more complex. 
   A possible solution is that the stellar populations of NGC\,1866 are coeval in close analogy with NGC\,1856. 
As demonstrated in this paper, the recent rotating models (e.g.\,Ekstr{\"o}m et al.\,2013; Dotter 2016) make a superb work in reproducing the main features of the observed CMDs. Nevertheless the way rotation manifests itself in real stars may be not fully captured in the rotating models.
   If we wish to properly investigate this possibility we must investigate much more deeply the evolution of rotating stars in the braking phase, and examine clusters of different ages, to explain the CMDs of these clusters in a coherent scheme. We postpone this analysis to a work in preparation by our group  (D'Antona et al.\,in preparation).
 
\section*{acknowledgments} 
\small 
 We thank the anonymous referee for several suggestions that have improved the quality of this manuscript.
APM, AFM and HJ acknowledge support by the Australian Research Council through Discovery Early Career Researcher Awards DE150101816 and DE160100851 and Discovery project DP150100862.

\bibliographystyle{aa}

\end{document}